\documentclass[ 12pt
  ]
  {article}

\usepackage{hyperref}
\usepackage[pdftex]{graphicx}

\def\be{\begin{equation}}
\def\ee{\end{equation}}
\def\bea{\begin{eqnarray}}
\def\eea{\end{eqnarray}}
\def\ahalf{{\textstyle{1\over2}}}

\newcommand{\bflambda}{\mbox{\boldmath$\lambda$\unboldmath}}
\newcommand{\bfbigpsi}{\mbox{\boldmath$\Psi$\unboldmath}}

\def\ie{{\it i.e.\,}}
\def\etal{{\it et al.   }}

\def\vcal{\mbox{$\cal V\,$}}
\def\ncal{\mbox{$\cal N\,$}}
\def\<{\langle}
\def\>{\rangle}

\begin{document}

\title{\textbf{Conformal Mapping and Bound States in Bent Waveguides}}

\author{E. Sadurn\'i and W. P. Schleich}

\date{Institut f\"ur Quantenphysik, Ulm Universit\"at, Albert-Einstein Allee 11 89081 Ulm - Germany.}

\maketitle

\begin{abstract}
Is it possible to trap a quantum particle in an open geometry? In this work we deal with the boundary value problem of the stationary Schroedinger (or Helmholtz) equation within a waveguide with straight segments and a rectangular bending. The problem can be reduced to a one dimensional matrix Schroedinger equation using two descriptions: oblique modes and conformal coordinates. We use a corner-corrected WKB formalism to find the energies of the one-dimensional problem. It is shown that the presence of bound states is an effect due to the boundary alone, with no classical counterpart for this geometry. The conformal description proves to be simpler, as the coupling of transversal modes is not essential in this case.\\


\end{abstract}

\maketitle


\section{Introduction}

The use of symmetry methods to understand bound states has an ancient origin and has given many of its fruits in atomic and subatomic physics - The work of Marcos Moshinsky is an excellent example of this \cite{mosh}. Over the years, however, other approaches to understand more complicated systems have given up symmetry in both classical and quantum mechanics \cite{stoeckmann}, \cite{luna}. Here we present one of the simplest ways of destroying continuous symmetries with purely quantum-mechanical consequences: A corner. We will show that a sharply bent waveguide supports a finite number of bound states whose nature has no classical counterpart. Due to the geometry, the traditional methods of integrability cannot be used to obtain the corresponding solutions.

This type of systems has been studied in the past, either with mathematical tools \cite{goldstone, exner1, exner2, exner3} proving the existence of bound states or by employing numerical methods for computing the corresponding energies and eigenfunctions \cite{exner1, schult, carini}. The approach in our considerations will be completely analytical, albeit we use approximations to perform computations. We want to show two different approaches to the problem in order to exhibit an apparent complexity of the description given in other studies, mostly numerical \cite{exnerl}. First we shall deal with a convenient basis of states which shows the binding as a consequence of the presence of corners: A sharp obstacle diffracts waves of an arbitrary wavelength. Some limitations of this method, such as the coupling of an infinite number of modes, will be pointed out. Finally, the use of conformal coordinates will be introduced as a natural description to find all the bound states for arbitrary bending angles.

We would like to emphasize that interesting applications may come in the form of bent wave guides carved in crystals, as suggested in \cite{opn}, using methods connected to the evolution of Airy profiles - a recent example can be found in \cite{paper by WPS}. The landscape looks promising, considering the many applications of controlling, bending and focusing waves.

\section{The existence of bound states}

\subsection{The oblique modes and the effective potential}

We show how to obtain a one dimensional Schroedinger equation with an effective potential along the longitudinal direction of the guide.
One of the challenges for describing our system is to find a suitable set of states. A natural way to introduce box eigenmodes in our system is by using oblique modes, as shown in the figure.

\begin{figure}[h!]
\begin{center}
\includegraphics[scale=0.4]{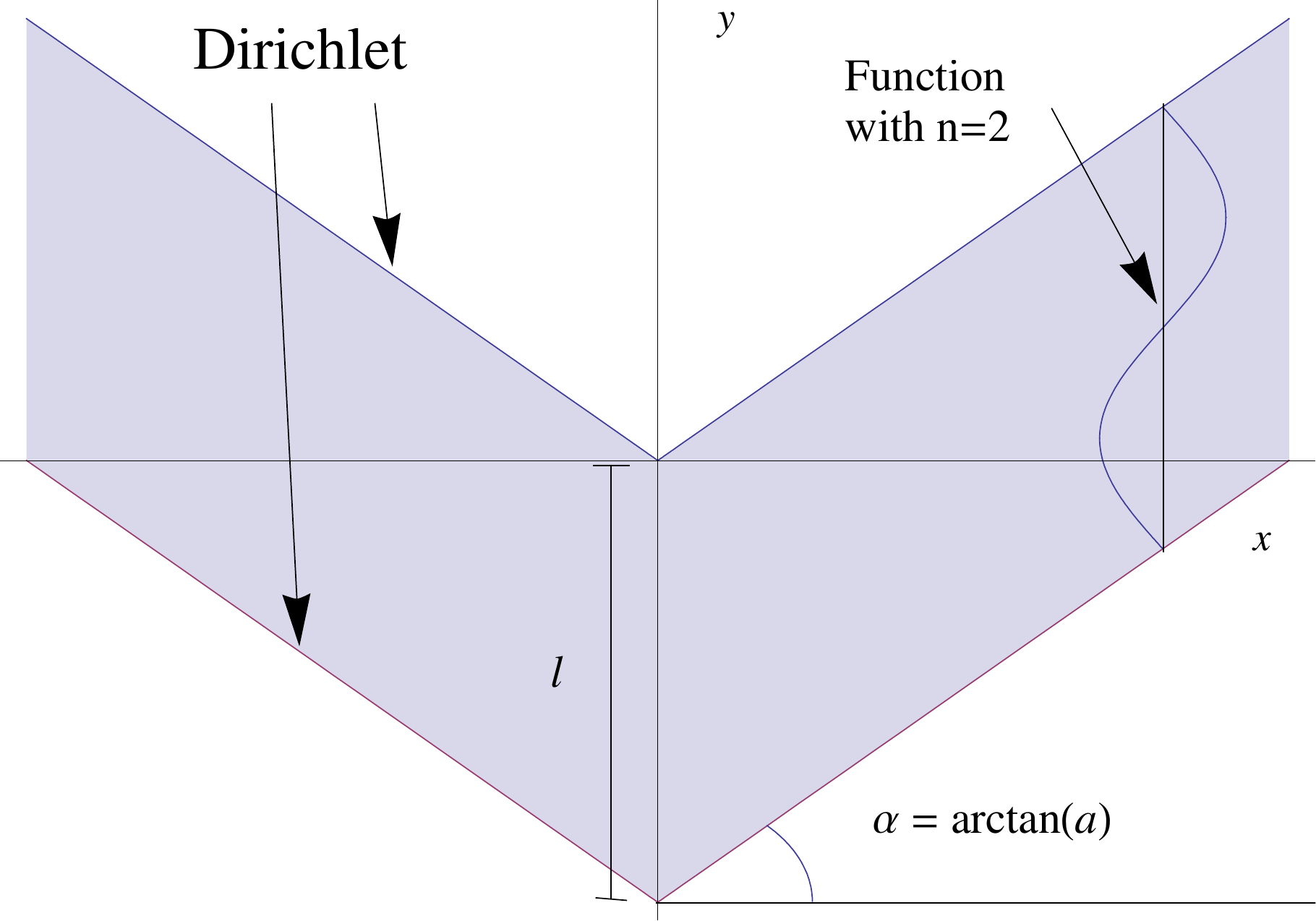}
\end{center}
\label{fig1}
\caption{Description of our system. The distance between corners is denoted by $l$. The slope $a$ is related to the bending angle $\alpha$. We show an oblique mode with $n=2$}
\end{figure}

Let us take units such that $\hbar^2/2m=1$ with $m$ the mass of the particle and write the stationary Schroedinger equation in the form

\bea
\nabla^2_{x,y} \psi + E \psi = 0
\label{1}
\eea
For the configuration in figure \ref{fig1} with Dirichlet conditions, we express the solution in the oblique basis as

\bea
\psi(x,y) = \sum_{n=1}^{\infty} \psi_n(x) \sin \left( \frac{n \pi}{l}(y-a|x|+l) \right)
\label{2}
\eea
and $\psi(x,y)\neq 0$ for $a|x|-l<y<a|x|$. The matrix elements of the Schroedinger operator in this basis can be obtained by using the expansion (\ref{1}) and integrating over the $y$ variable. Let us denote the vector wave function by $\bfbigpsi(x)$. Using the following definition for matrix elements

\bea
\mathbf{N}_{nm} = \delta_{nm} n, \qquad \mathbf{M}_{mn} = (1-(-)^{m+n}) \frac{mn}{m^2-n^2} 
\label{7}
\eea
and employing a gauge transformation given by

\bea
\tilde{\bfbigpsi}(x) = \exp \left( -\frac{2a|x|}{l} \mathbf{M} \right) \bfbigpsi(x) \\
\tilde{\mathbf{N}}(x) = \exp \left( -\frac{2a|x|}{l} \mathbf{M} \right) \mathbf{N} \exp \left( \frac{2a|x|}{l} \mathbf{M} \right)
\label{8}
\eea
we reduce the Schroedinger equation to

\bea
\left[\frac{d^2}{dx^2} - \left( \frac{\pi}{l} \right)^2 (a^2+1) \tilde{\mathbf{N}}(x)^2 - \left(\frac{2a}{l}\right)^2 \mathbf{M}^2 + E  \right] \tilde{\bfbigpsi}(x)=0
\label{9}
\eea
where we see an effective potential of the form

\bea
V_{eff} = \left( \frac{\pi}{l} \right)^2 (a^2+1) \tilde{\mathbf{N}}(x)^2 + \left(\frac{2a}{l}\right)^2 \mathbf{M}^2.
\label{10}
\eea
This matrix couples, in principle, all components of $\tilde{\bfbigpsi}$. The origin of $\mathbf{M}$ and $\mathbf{N}^2$ is related to the momentum and the energy of a particle in a box. The $x$ dependence of the potential $V_{eff}$ comes completely from the non-commutability of $\mathbf{N}$ and $\mathbf{M}$ and is non-differentiable at the corner due to $|x|$. Another important feature is that a change of units  $x' = x/l, E'=E l^2 $ in equation (\ref{9}) leads to a scale-free equation, thus proving that the energy $E$ must scale as $l^{-2}$. Therefore, the presence of bound states in this problem is independent of the scale. This is in contrast with other bent systems with smooth boundaries \cite{exner1, exner2, exner3}.

\subsection{Solutions from corner-corrected WKB}

In previous works by Wheeler, Bestle and Schleich \cite{anti}, the WKB approximation was corrected in the presence of potentials with corners. As a result, the Schroedinger equation with a potential could be approximated by a WKB equation with a source term. By neglecting completely the effect of the potential except at the non-differentiable point, the WKB equation with a source could be interpreted as a free Schroedinger equation with a $\delta$ potential. In our case, this would cause a binding effect depending on the energy-dependent amplitude of the $\delta$. We shall apply this idea in order to cope with $V_{eff}$ obtained in the last section, with the additional feature that $V_{eff}$ is a matrix potential. 

\begin{figure}[h!]
\begin{center}
\includegraphics[scale=0.6]{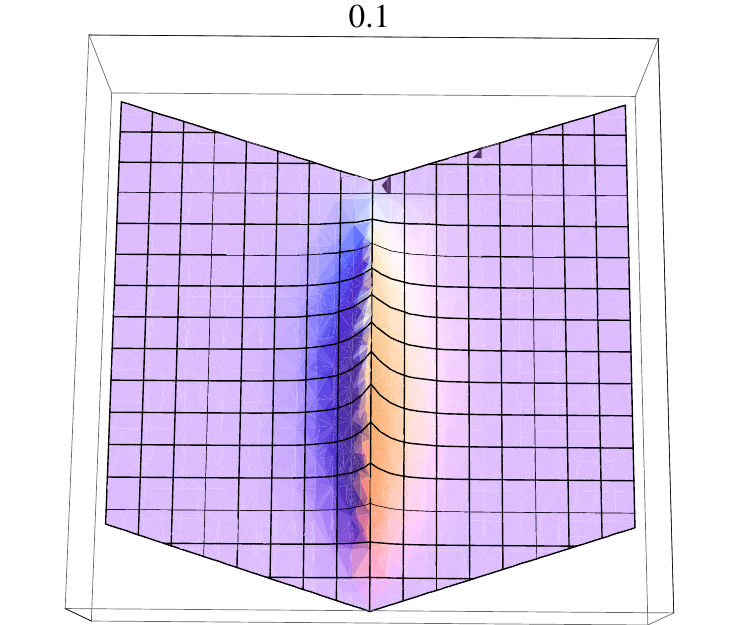}
\includegraphics[scale=0.6]{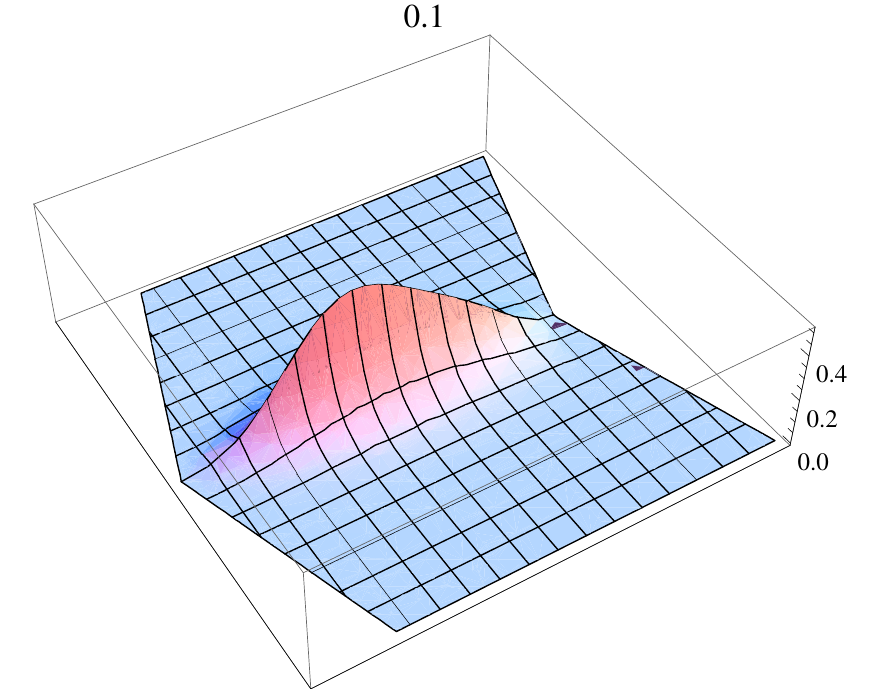}
\end{center}
\caption{Ground state wavefunction for $a=0.1$. A peak in the relevant region is visible. The decay with distance is exponential}
\label{fig2}
\end{figure}
\begin{figure}[h!]
\begin{center}
\includegraphics[scale=0.6]{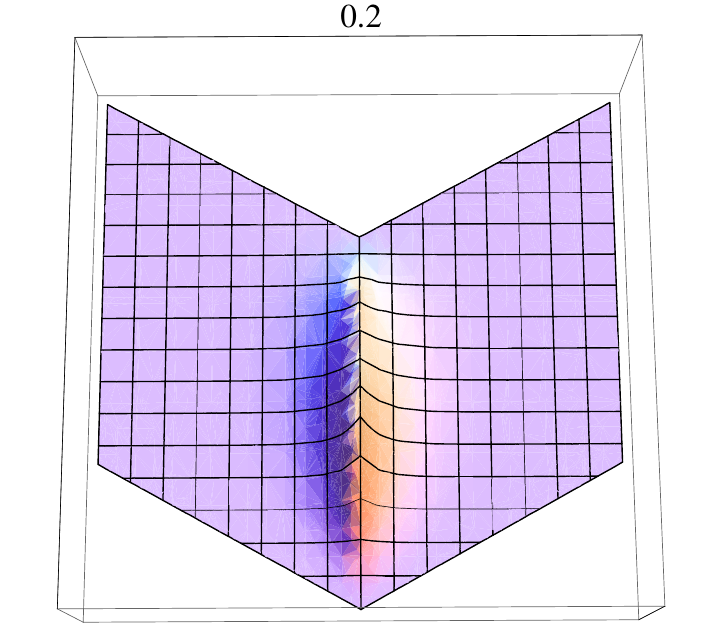}
\includegraphics[scale=0.55]{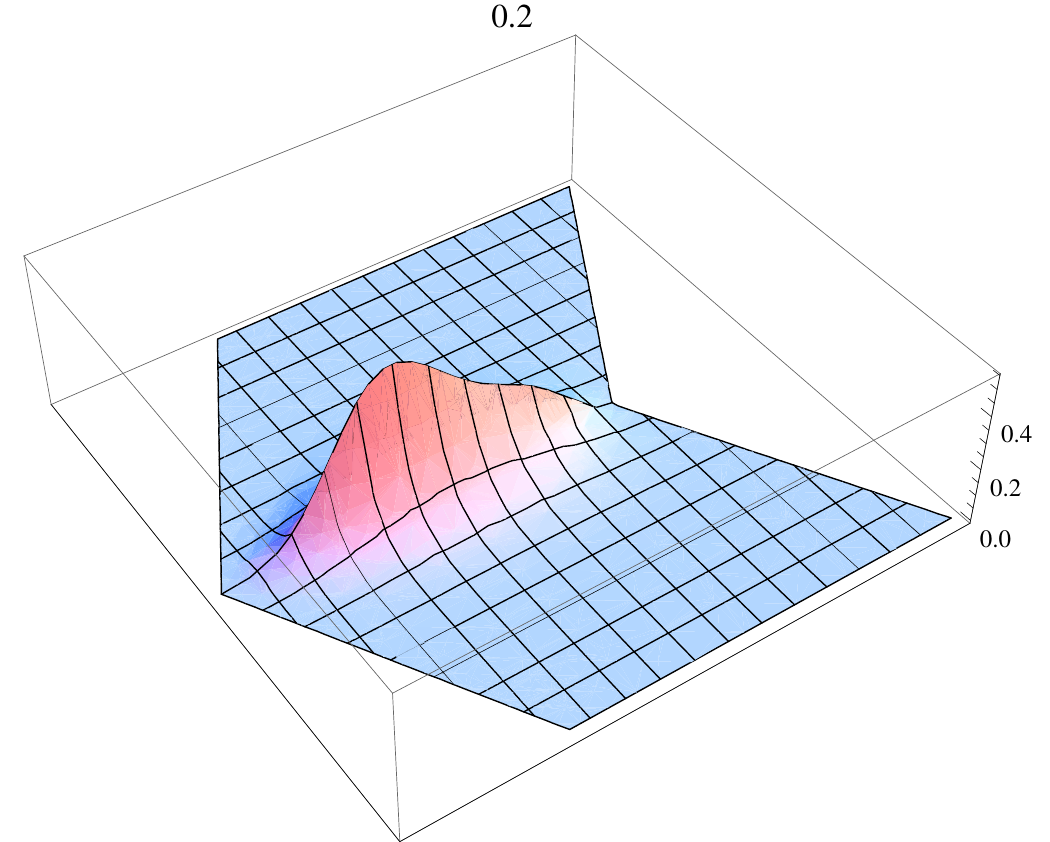}
\end{center}
\caption{Ground state wavefunction for $a=0.2$. The peak becomes more pronounced as $a$ increases. For large values of $a$, the description is no longer valid.}
\label{fig2bis}
\end{figure}

We start with the Schroedinger equation

\bea
\frac{d^2 \mathbf{u}(x)}{dx^2} + \bflambda^{-2}(x) \mathbf{u}(x) = 0
\label{1.1}
\eea
where $\bflambda = \left(E-V_{eff} \right)^{-1/2}$ is now a matrix. The corner-corrected WKB formalism given in \cite{anti} can be easily generalized to matrix potentials. A WKB scattering equation (such as eq (13) in \cite{anti}) can be obtained for (\ref{1.1}). The resulting equation, in turn, can be treated by keeping first order terms in $a$ and considering a negligible potential except at $x=0$. This gives the isolated effect of the corner in the form

\bea
\frac{d^2 \mathbf{u}(x)}{dx^2} + \left( E-\frac{\pi^2}{l^2} \mathbf{N}^2 \right)\mathbf{u}(x) = \vcal_{sc} \mathbf{u}(x) \nonumber \\
\vcal_{sc} \simeq -\frac{a \pi^2}{l^3} [\mathbf{N}^2,\mathbf{M}] \left( E- \frac{\pi^2}{l^2} \mathbf{N}^2 \right)^{-1} \delta(x).
\label{1.12}
\eea
The energies for bound states in this problem are obtained by i) imposing the appropriate boundary conditions $\mathbf{u}(x)=0$ at $x=\pm \infty$, ii) substituting the corresponding solutions in the condition for the jump in the derivative (due to $\delta$), namely

\bea
\frac{d \mathbf{u}}{dx}\vert_{0+} - \frac{d \mathbf{u}}{dx}\vert_{0-} = -\frac{a \pi^2}{l^3} [\mathbf{N}^2,\mathbf{M}] \left( E-\frac{\pi^2}{l^2} \mathbf{N}^2 \right)^{-1} \mathbf{u}_0.
\label{1.13}
\eea
The solutions are found to be

\bea
\mathbf{u}(x)= \ncal \left( \frac{\pi^2}{l^2}  \mathbf{N}^2 - E \right)^{-1/4} \exp \left( - |x| \sqrt{ \frac{\pi^2}{l^2} \mathbf{N}^2 - E} \right) \mathbf{u}_0
\label{1.14}
\eea
with $\ncal$ a normalization factor. The vector $\mathbf{u}_0$ is determined by the condition coming from the jump in the derivative (\ref{1.13}), namely

\bea
\left[ \frac{a \pi^2}{l^3}  [\mathbf{N}^2,\mathbf{M}] + 2 \left( \frac{\pi^2}{l^2} \mathbf{N}^2 - E \right)^{3/2} \right] \left[ E - \frac{\pi^2}{l^2} \mathbf{N}^2 \right]^{-1} \mathbf{u}_0 = 0
\label{1.15}
\eea
This is a linear equation for $\mathbf{u}_0$ and the wave function can be obtained by choosing $\mathbf{u}_0$ as a member of the null space of the operator in (\ref{1.15}). The scaled energies $\tilde E = (l/\pi)^2 E $ are obtained from (\ref{1.15}) by using the determinant

\bea
\det \left( \frac{a}{\pi} [\mathbf{N}^2,\mathbf{M}] + 2 \left[ \mathbf{N}^2 - \tilde E  \right]^{3/2} \right) = 0
\label{1.16}
\eea
The $2\times2$ case can be solved easily by inserting in (\ref{1.16}) the diagonal matrix $\left( \mathbf{N}^2-\tilde E \right)^{3/2}= \mbox{diag} \left( \left(1-\tilde E \right)^{3/2}, \left(4-\tilde E \right)^{3/2} \right) $. The resulting energy for the ground state is

\bea
E_0 = \left(\frac{\pi}{l} \right)^2 \left( \frac{5}{2} - \frac{3}{2} \sqrt{1+\left( \frac{2}{3} \right)^2 \left( \frac{2a}{\pi} \right)^{4/3}} \right)
\label{1.18}
\eea
This energy is below the threshold $(\pi/l)^2$. The level $E_1$ can also be obtained; it is above $4(\pi/l)^2$. Therefore, in this $2\times2$ approximation, the only state which decays exponentially with the distance is the ground state. The wave function can be found straightforwardly once $u_0$ is obtained. We show two cases in figures \ref{fig2} \ref{fig2bis}.

The solutions thus obtained show that a bound state develops due to the corner alone, producing a $\delta$ potential and coupling the oblique modes - such coupling is essential, since a $1 \times 1$ approximation makes the interaction vanish. Yet, we may ask: where are the excited states below threshold due to an increasing value of $a$? In the following, we analyze this problem using conformal coordinates as a way to obtain a suitable basis. Such a basis cannot be obtained by means of perturbation theory in $a$ using the solutions of this section.

\section{The conformal map}

Our aim now is to show that a non-trivial boundary value problem such as the bent waveguide can be mapped smoothly to a straight guide with an effective potential (equivalently, a problem with position-dependent mass). The simplest way to describe this process is by using complex variables. Denote $z=x+iy$, $\xi=u+iv$ and consider a transformation such that $\xi=F(z)$. We impose the following conditions on $F$:

1) The image of the bent waveguide is an infinite straight strip described by $0 < u < F(c)$.

2) The Laplacian $\nabla^2_{xy}$ is transformed into a quadratic form without cross-terms (The matrix $\mathbf{M}$ will not appear).

\begin{center}
\begin{figure}
\begin{center}
\includegraphics[scale=0.45]{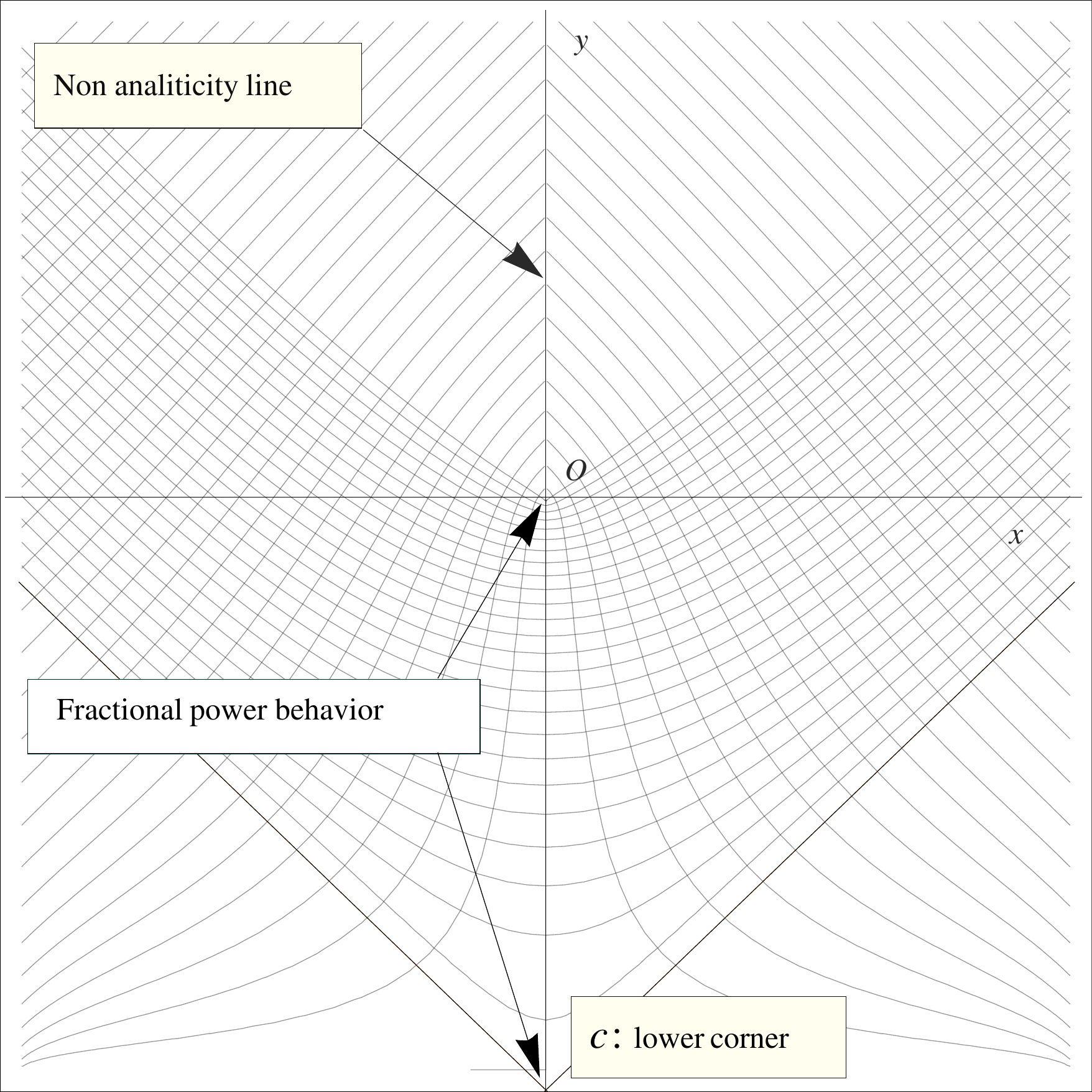}
\end{center}
\caption{Contour lines of the coordinates $u,v$. We denote the lower corner by $c$ and place the upper corner at the origin $O$, see the text.}
\label{map2}
\end{figure}
\end{center}

3) The transformation is smooth except at the points $O$ (upper corner) and $c$ (lower corner) in figure \ref{map2}.

The condition 1) allows the use of a complete set of functions $\lbrace \sin(nu/F(c))\rbrace$ for the Dirichlet problem and $\lbrace \cos((n+1/2)u/F(c))\rbrace$ for the Neumann problem, given that the contour lines meet the boundaries orthogonally. The condition 2) precludes the appearance of terms proportional to $\partial_u$: The integral $\int_{0}^{\pi F(c)} \sin(nu/F(c)) \partial_u \sin(mu/F(c))$ would produce unbounded matrix elements due to well-known properties of the momentum operator for a particle in a one-dimensional box. Requirement 3) ensures that any suspicious behavior of the transformation is due to the corners.

\subsection{Explicit transformation}

The requirements described above are fulfilled by choosing $F$ analytic for all the points in the interior of the waveguide. Orthogonality is reached by the Cauchy-Riemann conditions. We introduce the operators

\bea
\partial_{z} = \partial_x - i \partial_y, \qquad \partial_{z^*} = \partial_x + i \partial_y
\label{2.1}
\eea

\bea
\partial_{\xi} = \partial_u - i \partial_v, \qquad \partial_{\xi^*} = \partial_u + i \partial_v
\label{2.2}
\eea
obtaining thus $\nabla^2_{xy}= \partial_{z}\partial_{z^*}$, $\nabla^2_{uv}= \partial_{\xi}\partial_{\xi^*}$. Using the analytic function $\xi=F(z)$, the Laplacian operator is transformed as $\partial_{\xi}\partial_{\xi^*} = |F'(z)|^{-2} \partial_{z}\partial_{z^*}$. Using the Cauchy-Riemann conditions we have $|F'(z)|^2 = \partial(u,v)/\partial(x,y)$, \ie the Jacobian of the transformation as a function of $z$. Our boundary value problem with the Schroedinger equation becomes

\bea
\left( \partial_{\xi}\partial_{\xi^*} + |\frac{dz}{d \xi}|^{2} E \right) \psi(u,v) = 0, \qquad u \in [0,F(c)], \quad v \in \bf{R} 
\label{2.5}
\eea
We give an explicit example of our conformal map in the following. Consider the Schwarz-Christoffel (SC) transformation \cite{conformal} of the upper half-plane onto the semi-infinite strip. By using a composition of a SC map for a trigon with internal angles $\pi/2$ together with a SC map for a trigon with internal angles $ \alpha \pi$ and $(1-\alpha) \pi$, we arrive at the bijective transformation between one half of the bent waveguide and a semi-infinite straight strip. The expression results in

\bea
\xi = F(z) = \int_{0}^{\sin^2(z/2)} dx (1-x)^{-\alpha} x^{-\alpha} = \int_{0}^{z} d\phi (\tan(\phi/2))^q \nonumber \\
= \beta \left(\sin^2(z/2), \frac{1-q}{2},\frac{1+q}{2} \right)
\label{2.9}
\eea
where we have used the definition of the incomplete beta function \cite{gradshteyn} and we have defined $q=1-2\alpha$ as the complementary angle of $2\alpha$ in $\pi$ radians. Therefore $0<q<1$. In passing we note that $F(c)=\sec (\pi q/2) $. The Jacobian is given by $|\frac{dz}{d\xi}|^2 = |\cot z/2|^{2q}$ and one can approximate it in terms of $\xi$ (or $u,v$) for some cases of interest:

\bea
|\frac{dz}{d \xi}|^2=\cases{f(q)|\xi|^{-2q/(q+1)} & near the origin \cr g(q)|F(c)-\xi|^{2q/(1-q)} & near $F(c)$\cr 1+4q\cos u e^{-|v|}& away from the corners}
\label{2.14}
\eea
with $f(q)=((1+q)/2)^{-2q/(q+1)}$, $g(q)=(2(1-q))^{2q/(1-q)}$. It is important to note that as one moves away from the corner in the direction of the arms, the Jacobian (\ref{2.14}) becomes a hyperbolic cotangent of a real variable and approches to unity exponentially fast. Therefore, interactions vanish for points far from the corners. In connection with the existence of other maps, we stress that any other transformation constructed in this way will have the same behavior near the corners: The fractional power law of the map is completely determined by the angle of the guides and therefore by our physical system.

\subsection{Approximate solutions in conformal coordinates}

\begin{figure}
\begin{center}
\includegraphics[scale=0.7]{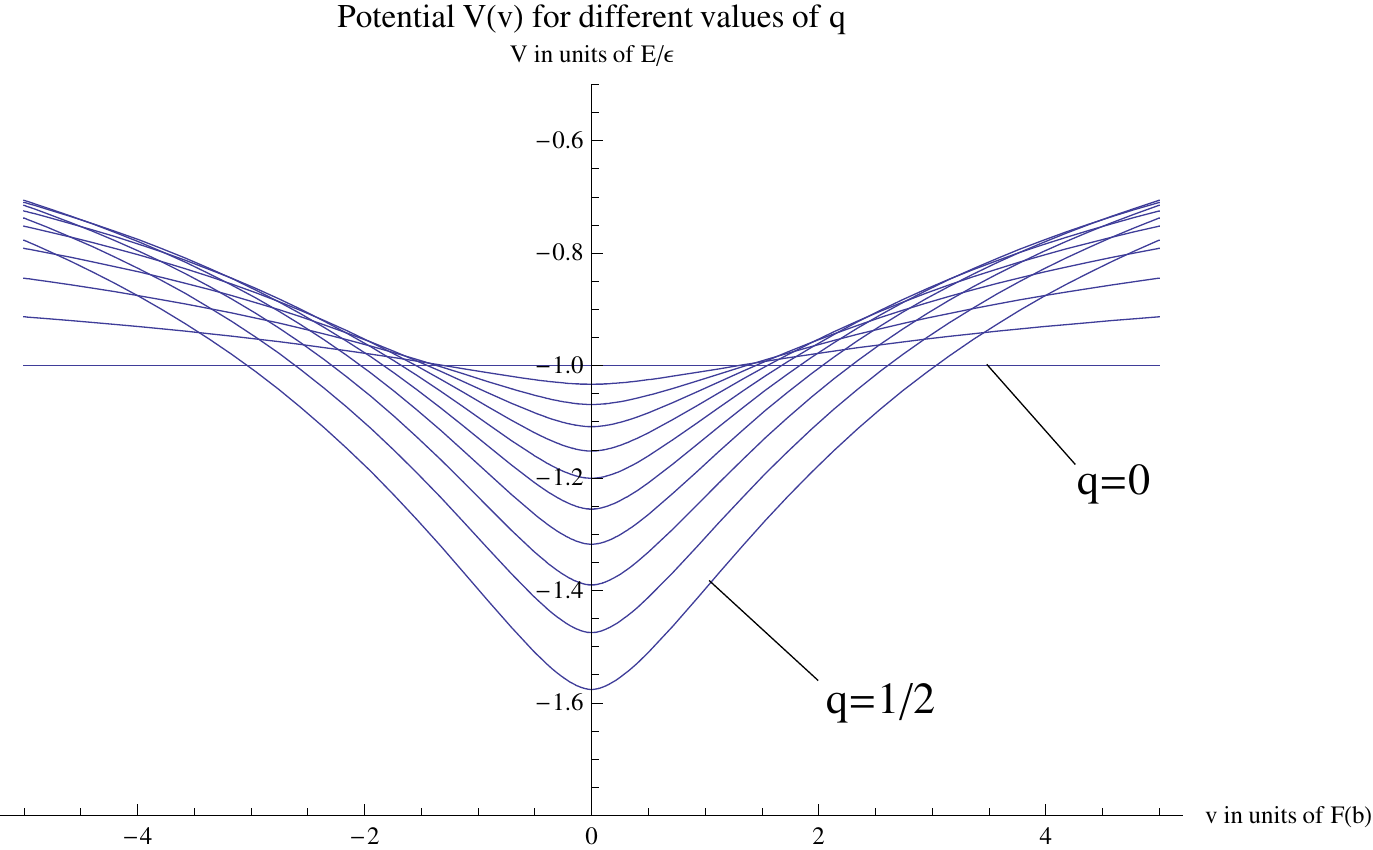}
\end{center}
\caption{A set of potentials for $n=1$, $q$ increases from $0$ to $1/2$ in $10$ steps. The transversal mode energy is shown at $-1$.}
\label{pot}
\end{figure}

\begin{figure}
\begin{center}
\includegraphics[scale=0.7]{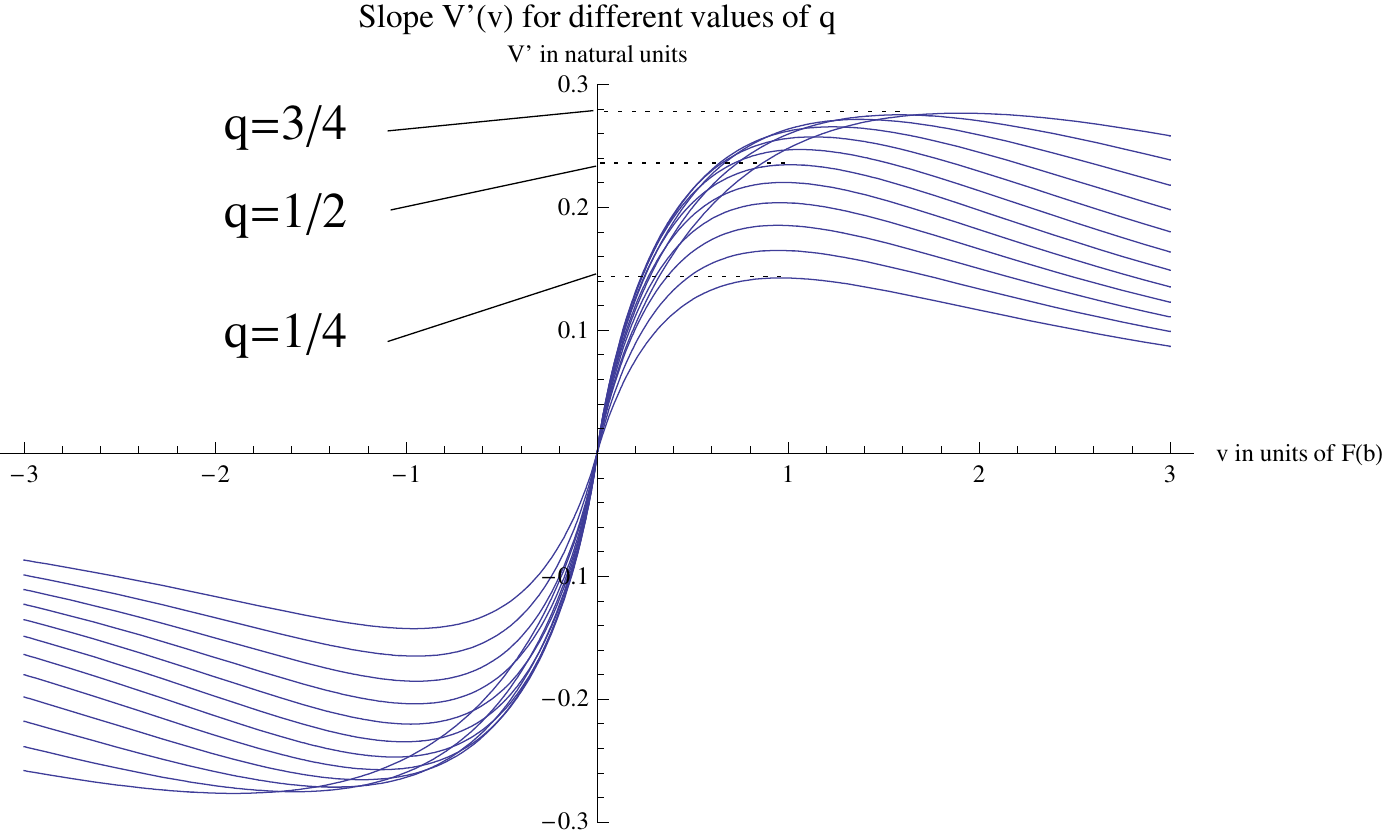}
\end{center}
\caption{The slope of the potentials as a function of $v$. Near the origin we see an abrupt change (a discontinuity appears in the limit $q=1$)}
\label{slope}
\end{figure}

One can try many methods of solution for the transformed boundary value problem given above. Here we give an example based on the fact that our careful construction allows certain approximations. Let us reduce (\ref{2.5}) to a one-dimensional problem by defining $\epsilon_n \equiv -n^2/F(c)^2$ and

\bea
V_{nm}(v) \equiv \< n | |\frac{dz}{d \xi}|^2 | m \> = \frac{2 F(c)}{\pi} \int_{0}^{\pi F(c)} du \sin \left( \frac{nu}{F(c)} \right) \sin \left( \frac{mu}{F(c)} \right) |\frac{dz}{d \xi}|^2
\label{2.6}
\eea
The set $\sqrt{2 F(c) / \pi} \sin(nu/F(c)) $ is obviously orthonormal. After multiplying (\ref{2.5}) by our complete set of functions and performing the corresponding integrals over $u$, we find that (\ref{2.5}) is equivalent to

\bea
\sum_{m} \left(- \delta_{nm} \partial_v^2 - E V_{nm}(v) \right) \psi_m(v) = \epsilon_n \psi_n(v)
\label{2.8}
\eea
Near the origin, our potential (\ref{2.6}) becomes
 
\bea
V_{nm}(v)= \frac{2 f(q)}{\pi F(c)} \int_{0}^{\pi F(c)} du 
\sin \left( \frac{nu}{F(c)} \right) \sin \left( \frac{mu}{F(c)} \right){\left( u^2 + v^2 \right)^{q/(q+1)}}
\label{2.15}
\eea
It is justified to approximate the potential by its diagonal entries $V_n(v)$, since $V_{nm}(v)$ decreases as we move away from the diagonal for fixed $v$. This can be seen by noting that for $n \neq m$ the integrand can take negative values and oscillates rapidly as $n-m$ increases. We have

\bea
\left(-\partial^2_v -E V_n(v) \right) \psi_n(v) = \epsilon_n \psi_n(v).
\label{2.16}
\eea
We include some plots of $V_n(v)$ and its derivative for different angles, figures \ref{pot}, \ref{slope}.

The potentials shown in the figure have zero derivative at $v=0$ (except for $q=1$) but, apart from that, their behavior is that of a shallow potential well whose derivative suffers an abrupt change in a small region near the corner. In the plot showing the slopes, we see that there is almost a jump in $V'_n(v)$ around $v=0$. The formalism developed by Wheeler, Schleich and Bestle \cite{anti} in order to deal with 'kinks' should be helpful here again. Although the only case in which the derivative is strictly discontinuous corresponds to $q=1$, the strong variation of $V_n'(v)$ occurs in a region which looks effectively like a corner for long wavelengths (we expect this approximation to work at low energies). As the simplest approximation, let us consider the corner-corrected WKB formula for quantization of bound states, namely

\bea
S = \sqrt{2} \int_{0}^{v_0} dv \sqrt{\epsilon_n+E V_n(v)}, \qquad \mbox{Classical orbit action}
\label{2.17}
\eea

\bea
4S=2(k+\ahalf)\pi - 4 \arctan (\frac{\Delta}{4}), \qquad \mbox{Phase shift}
\label{2.18}
\eea

\bea
\Delta = E \left( V'_n(0+) - V'_n(0-) \right) \left( 2(\epsilon_n+E V_n(0)) \right)^{-3/2}, \qquad \mbox{Jump in the slope}
\label{2.19}
\eea

\begin{figure}[t!]
\begin{center}
\includegraphics[scale=0.5]{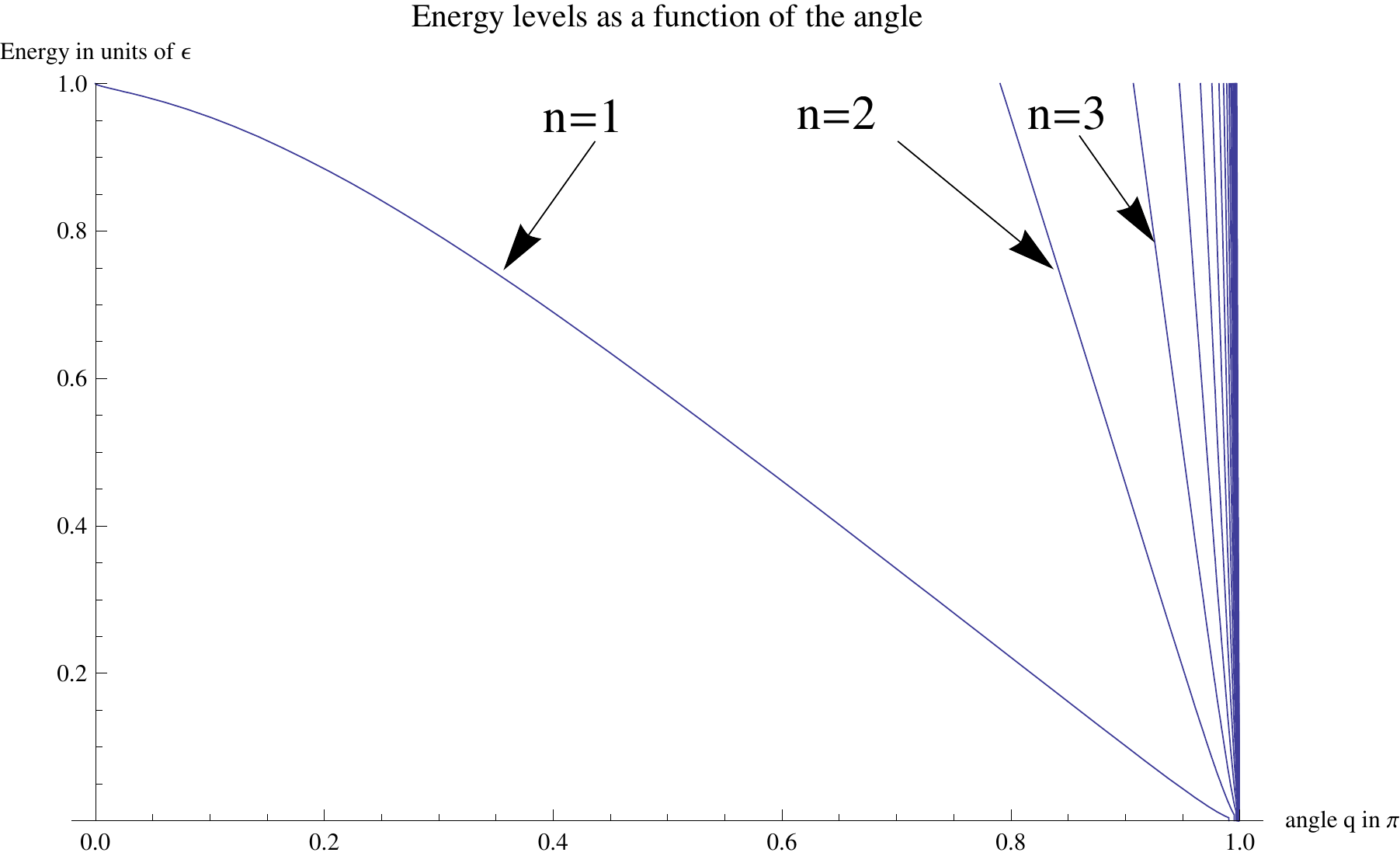}
\end{center}
\caption{Energies as a function of the angle. There is a bound state for all $q>0$ as a consequence of the corner. An unlimited number of bound states appear as $q$ increases.}
\label{final}
\end{figure}

Remarkably, one can obtain good results by neglecting the classical orbit $S=0$ and keeping the corner effect alone. Using (\ref{2.17}), the approximation implies

\bea
\frac{\Delta}{4} = \tan \left( \frac{2k+1}{4} \pi \right) = 1, \qquad k \quad \rm{even}
\label{2.20}
\eea
finding thereby that all even states obey this relation, but the excited states are not supported by such a weak potential. The excitations come from the solutions of (\ref{2.20}) by varying $n$ and not $k$. This results in a transversal node structure, instead of longitudinal. With these considerations, the solution to our problem (\ref{2.20}) requires $V_n(0)$ and $V_n'(0)$ as the only input. Such quantities can be estimated by means of (\ref{2.15}) for $m=n$. As mentioned before, we replace the jump in the derivative 
$ V'_n(0+) - V'_n(0-)$ by twice the maximum of $V'_n(v)$ around the origin (see the figure). We denote such quantity as $V'_{max}$. Squaring (\ref{2.20}) we obtain a cubic relation for $E$ :

\bea
\left( n^2 \cos^2 (\pi q/2) - E V_n(0) \right)^3 = \frac{1}{32} E^2 (V'_{max})^2
\label{2.22}
\eea
One can estimate $V_n(0)$ and $V'_{max}$ as

\bea
V_n(0) \simeq \left( \cos (\pi q/2) \right)^{2q/(q+1)}, \quad V'_{max} \simeq \frac{1}{4} q \left(\cos (\pi q/2) \right)^{(2q+1)/(q+1)}.
\label{2.24}
\eea
With this, the energies in (\ref{2.22}) can be expressed in terms of $n$ and $q$ using Cardano's formula. We present the results graphically in figure \ref{final}.

\begin{figure}[t!]
\begin{center}
\includegraphics[scale=0.45]{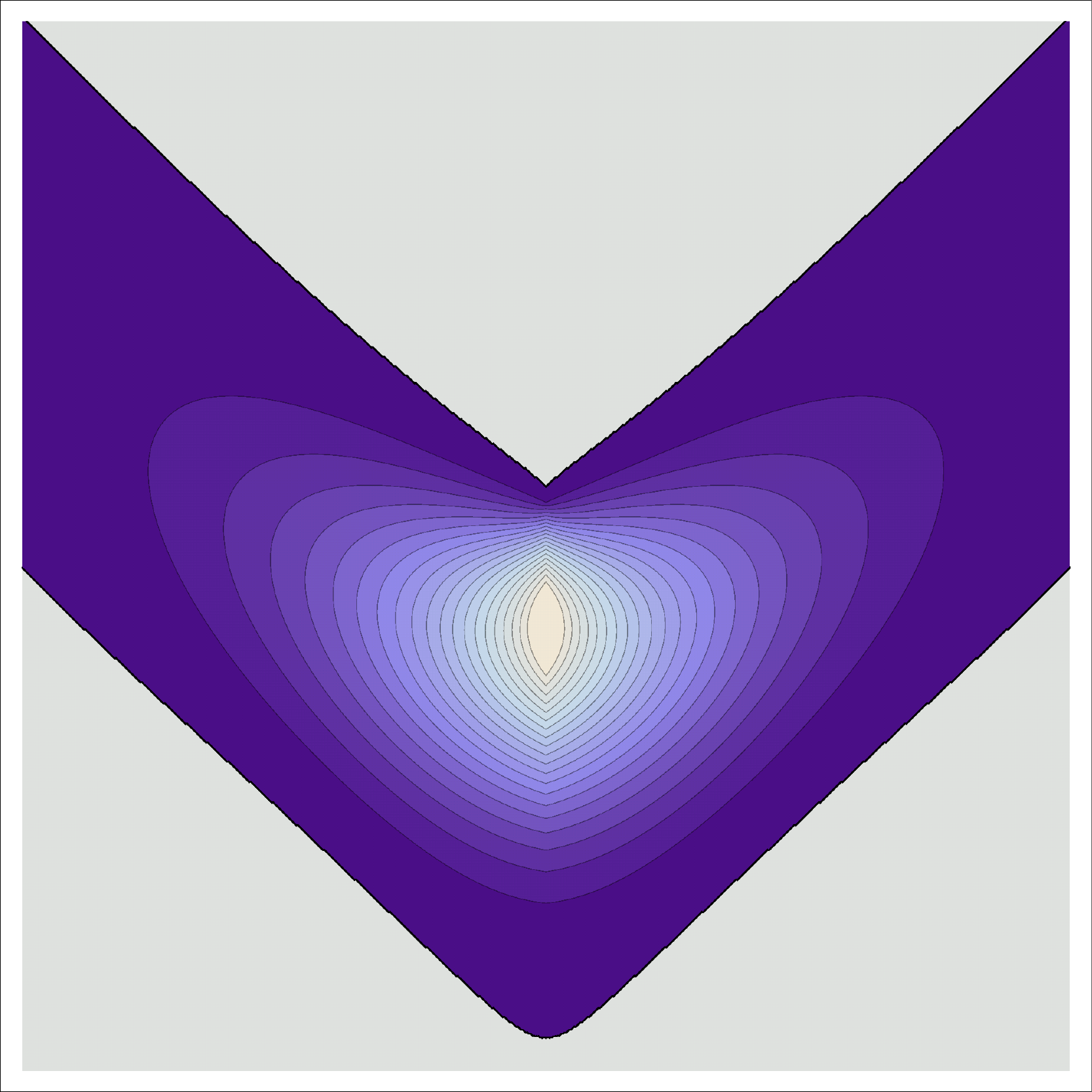}
\end{center}
\caption{Contourplot of the probability density for $q=0.5$ and $n=1$ (ground state energy $E \simeq 0.56$). The peak of the distribution is located near the upper corner, as expected. This is achieved with only one transversal mode in conformal coordinates}
\label{density}
\end{figure}

The agreement of the estimated solutions with previous numerical results \cite{carini, dietz} is qualitatively good, considering that the action has been completely neglected in favor of the phase shift. Also, the off-diagonal elements of the matrix potential (\ref{2.15}) have been ignored. The trend of the curves as functions of the bending angle is comparable with the numerical results obtained in \cite{carini} for several modes: the energies decrease and more states appear below threshold as $q \mapsto 1$. The node structure of our wavefunctions is also compatible with the solutions obtained in \cite{carini} both numerically and experimentally. Remarkably, the fractional power behavior of the Jacobian alone is responsible for the structure of the spectrum of this system. We include a plot of the resulting wave density for the ground state of an L-shaped guide.

In what touches the corrections, we expect the following: The action appears to be important as $q$ increases (deeper wells). Therefore (\ref{2.22}) receives corrections which increase the energy. As a consequence, we expect a shift of the critical angles towards the value $q=1$. Precisely at this value of $q$ we also expect a non-zero lower bound for the energy $E_{min} \sim 1/4$, as can be seen by identifying such an extreme case with a guide of double thickness.

\section{Discussion}

In the first part we have shown that the existence of bound states could be understood as a corner effect and the coupling of transversal modes. We could learn that diffraction was essential to the final result, but the coupling of the modes gave an extra complication. The second part, however, shows that an appropriate set of coordinates may keep the effect of the corner without the coupling of such modes. This makes contact with recent dicussions \cite{teufel} in which perturbation theory seems doomed in the presence of unbounded curvature - our example fits perfectly in such class of problems. One can follow either approach to solve the problem, but its apparent complexity - e.g. the many modes used in \cite{exnerl} for the L-shape - dissapears using the second method. At the end we can conclude that a simple way of breaking a continuous symmetry such as a corner still allows for a simple description as long as the important information is used: diffraction and conformality.

The next task is to find the correction to the energy levels due to the classical orbits entering through the action. But we can say that for now our goal has been reached. As a final comment, the problem of bound states in multiple junctions -e.g. X-shaped and $*$-shaped geometries - lends itself for this kind of treatment by using Neumann boundary conditions. Due to the properties of our conformal coordinate system, this can be achieved by replacing $\sin(nu/F(c))$ by $\cos((n+\ahalf)u/F(c))$ in our discussion.


\section*{Acknowledgments}
E. Sadurn\'i is grateful to the organizers of the Symposium {\it Symmetries in Nature\ } for their kind hospitality.

\end{document}